\documentclass[aps,prd,twocolumn,10pt,superscriptaddress,nofootinbib]{revtex4-2}

\usepackage{amsmath,amssymb,amsfonts}
\usepackage{braket}
\usepackage[dvipsnames]{xcolor}
\usepackage{graphicx}
\usepackage[pdftex,breaklinks,colorlinks,
linkcolor=Blue,
citecolor=magenta,
anchorcolor=red,
urlcolor=cyan,
pdfencoding=auto]{hyperref}
\usepackage{orcidlink}
\hypersetup{colorlinks=true}

\begin{document}

\title{Topology sums, sectorwise holography, and horizon normalcy}

\author{Naman Kumar\,\orcidlink{0000-0001-8593-1282}}
\affiliation{Department of Physics, Indian Institute of Technology Gandhinagar, Palaj, Gujarat, 382355, India}
\email{namankumar5954@gmail.com, naman.kumar@iitgn.ac.in}
\date{\today}

\begin{abstract}
The ``holography of information'' (HoI) principle argues that gravitational
theories can encode information redundantly in asymptotic observables. HoI is
ultimately a nonperturbative statement, but its standard motivation often uses
semiclassical gravitational constraints, the boundary nature of the
Hamiltonian, and vacuum-sector cyclicity. We ask what becomes of this
argument when the same semiclassical path-integral reasoning allows topology
sums that generate baby-universe or $\alpha$-sector data. We do not assume
that such sectors survive in every unitary completion: the Baby Universe
Hypothesis of McNamara and Vafa instead suggests that in consistent
$d>3$ quantum gravity the baby-universe Hilbert space is one-dimensional, and
standard AdS/CFT may realize this branch. Our analysis is therefore
conditional. If the baby-universe Hilbert space is nontrivial, as in the
Marolf--Maxfield formulation of topology-summed path integrals and in
ensemble-like examples such as JT gravity, then the HoI statement is naturally
refined to an $\alpha$-sectorwise form,
\(
\overline{\mathcal A_\infty^{(\alpha)}|0_\alpha\rangle}
=
\mathcal H_\alpha ,
\)
rather than a completeness statement on the full topology-summed Hilbert
space. In a fixed $\alpha$-sector, HoI may obstruct the AMPS factorization and
allow a smooth horizon; in an unconditioned topology-summed state, the
sector-independent obstruction to factorization is not automatic. A Bell-pair
diagnostic shows that a sector-independent smooth interior requires the
sector-dependent interior reconstructions to be aligned, or the sector label
to be accessible. Thus the HoI-based absence of firewalls becomes conditional
on global sector data, in tension with the generally covariant expectation emphasized by Bousso that horizon normalcy should be a state--independent property of local semiclassical geometry. This remains true unless additional input removes, resolves, or makes
accessible the $\alpha$-sectors. Conversely, if the exact theory collapses
$\mathcal H_{\rm BU}$ to one dimension, the obstruction discussed here is
absent.
\end{abstract}

\maketitle

\section{Introduction}

The holography of information (HoI) principle
\cite{Laddha:2020kvp,Raju:2021lwh,Geng:2026asi} proposes that gravitational
theories store information differently from ordinary local quantum field
theories. Information that semiclassical effective field theory would assign
to the bulk interior of a Cauchy slice is argued to be redundantly available
in asymptotic observables. This idea is especially relevant for black hole
physics, because it challenges the factorization assumption used in the AMPS
paradox \cite{Almheiri:2012rt}. If the asymptotic algebra is complete in the
relevant representation, then the black hole interior need not define an
independent tensor factor inaccessible from infinity.

A sharp asymptotically flat formulation of this idea was given by Laddha,
Prabhu, Raju and Shrivastava \cite{Laddha:2020kvp}. Their first result states
that any two distinct states in the Hilbert space of massless particles can
be distinguished by observables in an infinitesimal neighbourhood of the past
boundary of future null infinity, $I^+_-$. Their second result states that
information available near a cut of future null infinity is already available
near any earlier cut. These results are derived subject to explicit
assumptions: the vacua of the full theory can be identified near $I^+_-$,
operators near $I^+_-$ can map any vacuum to any other vacuum, and the full
Hamiltonian is bounded below. The second result additionally assumes that the
relevant null-infinity commutators and constraints receive only local
corrections in the full theory.

HoI is ultimately a nonperturbative statement. Semiclassical gravitational
constraints, the boundary Hamiltonian, and vacuum-sector cyclicity support it
in controlled backgrounds and representations, but do not establish
nonlocal information encoding for arbitrary semiclassical states. In
particular, Ref.~\cite{Geng:2025fate} exhibits both settings in which such
encoding is detectable and settings in which the relevant protocols are
suppressed and perturbatively localized observers emerge. We therefore use
HoI conditionally: assuming it holds in a chosen representation, we ask
whether an additional topology-induced $\alpha$-decomposition obstructs its
extension to the full Hilbert space.

This paper examines whether this HoI argument, formulated within a vacuum
sector controlled by the asymptotic algebra, automatically extends to a
Hilbert space produced by a topology-summed gravitational path integral. We
will use the word ``sector'' in two related but distinct senses. First, the
ordinary HoI statement is already representation-wise: it is formulated after
choosing the relevant vacuum, asymptotic boundary conditions, and superselection
sector. Second, topology sums may introduce an additional baby-universe or
$\alpha$-sector decomposition. Our question concerns this second kind of
sector structure. Does HoI remain complete only after conditioning on
$\alpha$, or does the exact theory remove, resolve, or make accessible the
$\alpha$-data so that no additional obstruction remains?

The issue is simple. The HoI argument is naturally formulated in a sector with
a cyclic vacuum,
\begin{equation}
\overline{\mathcal A_\infty |0\rangle}
=
\mathcal H_{\rm vac},
\label{eq:sector-hoi-intro}
\end{equation}
where $\mathcal A_\infty$ denotes the asymptotic algebra. In this form, the
argument involves a vacuum projector within the chosen representation. By
contrast, topology-summed gravitational path integrals can lead, in some
interpretations, to baby-universe or $\alpha$-sector decompositions,
\begin{equation}
\mathcal H_{\rm phys}
=
\int^\oplus d\alpha\, \mathcal H_\alpha ,
\label{eq:direct-integral-intro}
\end{equation}
or to the corresponding direct-sum version. In such a situation, the natural
HoI statement is not a single cyclicity condition on
$\mathcal H_{\rm phys}$, but rather the $\alpha$-sectorwise statement
\begin{equation}
\overline{
\mathcal A_\infty^{(\alpha)} |0_\alpha\rangle
}
=
\mathcal H_\alpha .
\label{eq:sector-hoi}
\end{equation}
Thus, if the $\alpha$-sector data survive as genuine superselection or
ensemble data, completeness in each $\alpha$-sector does not imply
completeness on the full topology-summed Hilbert space without additional
input.

The possible nonperturbative interpretations form a trichotomy, presented in
more detail in Sec.~\ref{sec:topology-jtA}. First, the topology-summed path
integral may give a genuine baby-universe Hilbert space. In the
Marolf--Maxfield construction, boundary-insertion operators commute and are
simultaneously diagonalized by $\alpha$-states; fixed $\alpha$-states restore
factorization, while a general baby-universe state describes an ensemble
\cite{Marolf:2020xie}. Second, a complete unitary theory may impose global
constraints, exact cancellations, or a null-state quotient that collapses the
baby-universe Hilbert space. A sharp version of this possibility is the Baby
Universe Hypothesis of McNamara and Vafa, according to which
$\dim\mathcal H_{\rm BU}=1$ in consistent unitary quantum gravity in $d>3$
\cite{McNamara:2020uza}. 
Third, the topology-summed path integral may describe an effective or open
system, as in low-dimensional ensemble models or AdS coupled to a
nongravitational bath, rather than a single closed unitary theory. Different aspects of this distinction---including the gravitational Gauss-law structure, the role of a graviton mass, information encoding, operator dressing, and the observable algebra---have been developed in Refs.~\cite{Geng:2020fxl,Geng:2021hlu,Geng:2023qwm,Geng:2025mechanism,Geng:2025algebras,Geng:2025wethair,Geng:2025case}.
We do not claim this bath- or ensemble-induced modification of HoI as new;
these references delimit the effective/open-system branch from the central
$\alpha$-sector obstruction studied below. The question addressed here is
distinct: the asymptotic algebra may be complete within every individual
$\alpha$-sector, while an obstruction can still arise when one asks whether
these sectorwise statements combine into completeness on the unconditioned
topology-summed Hilbert space.

This distinction is important because the exclusion of non-factorizing
wormholes in standard AdS/CFT is naturally a top-down constraint on the
correctly defined bulk path integral. Starting from an exact CFT, factorization
and unitarity restrict which semiclassical configurations may contribute to
the AdS description \cite{Maldacena:1997re}. This may indeed eliminate
ordinary baby-universe $\alpha$-sectors even at the correctly defined
semiclassical level. But from a bottom-up semiclassical saddle expansion, such
an exclusion is not a consequence of the local equations of motion alone; it
requires global input in the definition of the path integral, such as a choice
of contour, cancellations among topologies, or a quotient by null states. The
same caution applies to HoI: promoting the semiclassical vacuum-sector
argument to a statement about the full nonperturbative Hilbert space also
requires knowing the correct representation and sector structure.

The consequence for black hole interiors is the main point of the paper. The
HoI argument is used to evade AMPS by denying that the interior and exterior
define independent factors in the exact theory. But that denial requires
asymptotic completeness in the relevant representation. We will argue that,
if topology sums survive as genuine baby-universe sector data, this statement
is weakened from completeness on the full topology-summed Hilbert space to
$\alpha$-sectorwise completeness. Consequently, the HoI-based avoidance of
AMPS becomes conditional on a choice of $\alpha$-sector, on an asymptotic
mechanism that resolves the sector label, or on a sector-independent
alignment of the interior reconstruction. Horizon normalcy then becomes
representation-dependent in the sense that its validity is not fixed solely
by the local semiclassical geometry near the horizon-crossing event. This
conditionality is in tension with the generally covariant expectation
emphasized by Bousso that horizon normalcy should be a state-independent
property of the local semiclassical geometry \cite{Bousso:2025udh}.

The structure of the paper is as follows. In Sec.~\ref{sec:hoi-sector} we
recall the representation-wise form of the HoI argument and the role of the
vacuum projector. In Sec.~\ref{sec:topology-jt} we motivate the
topology-summed Hilbert-space structure, present the trichotomy of possible
nonperturbative interpretations in Sec.~\ref{sec:topology-jtA}, use
topology-summed Jackiw--Teitelboim (JT) gravity as a diagnostic example, and
isolate the algebraic obstruction to a sector-independent HoI statement on a
nontrivial topology-summed Hilbert space. In Sec.~\ref{sec:normalcy} we show
how $\alpha$-sectorwise HoI makes HoI-based horizon normalcy
representation-dependent, with a minimal Bell-pair diagnostic quantifying the
obstruction. We conclude in Sec.~\ref{sec:discussion}.

\section{HoI and the role of a vacuum sector}
\label{sec:hoi-sector}

The abstract algebraic core of the HoI argument is most transparent in a
chosen representation with a cyclic vacuum. Let $\mathcal A_\infty$ denote the
asymptotic algebra and let $|0\rangle$ be the relevant vacuum. The cyclicity
condition in that representation is
\begin{equation}
\mathcal H_{\rm vac}
=
\overline{\mathcal A_\infty |0\rangle}.
\label{eq:cyclic}
\end{equation}
In ordinary quantum field theory, such cyclicity is closely related to the
Reeh--Schlieder theorem. In gravity, HoI combines this cyclicity with the
boundary nature of the Hamiltonian and the gravitational constraints.

The null-infinity argument of Ref.~\cite{Laddha:2020kvp} should be read in
this representation-wise sense. Its first result concerns the Hilbert space of massless
particles built from the vacua and radiative excitations. The assumptions
that vacua can be identified near $I^+_-$ and that the algebra near $I^+_-$
can map any vacuum to any other vacuum are precisely assumptions about the
asymptotic control of the vacuum sector. The boundedness of the full
Hamiltonian supplies the spectral input needed for the Reeh--Schlieder-type
argument.

The same structure appears in the canonical AdS version. The gravitational
Hamiltonian is an asymptotic observable,
\begin{equation}
H\in \mathcal A_\infty .
\end{equation}
If the relevant vacuum is the unique ground state in the sector, then one may
formally write
\begin{equation}
P_0
=
|0\rangle\langle 0|
=
\lim_{\beta\to\infty} e^{-\beta(H-E_0)} ,
\label{eq:vacuum-projector}
\end{equation}
where $E_0$ is the vacuum energy. Since $H$ belongs to the asymptotic algebra,
the vacuum projector is then regarded as an element of the same algebra,
\begin{equation}
P_0\in \mathcal A_\infty .
\end{equation}

Now let
\begin{equation}
|n\rangle=X_n|0\rangle,
\qquad
|m\rangle=X_m|0\rangle,
\qquad
X_n,X_m\in\mathcal A_\infty .
\end{equation}
Then the rank-one operator
\begin{equation}
Q_{nm}
=
|n\rangle\langle m|
=
X_n P_0 X_m^\dagger
\end{equation}
also belongs to $\mathcal A_\infty$. Linear combinations of such operators
generate the bounded operators on the sector. Thus, within the sector,
\begin{equation}
\mathcal A_\infty\simeq \mathcal B(\mathcal H_{\rm vac}) .
\label{eq:sector-complete}
\end{equation}

This is the sense in which the HoI argument is normally formulated within a
chosen representation. The important point for the present paper is not that
this reasoning is wrong, but that it assumes a representation with a
well-defined vacuum projector. If
the full gravitational Hilbert space has additional superselection data, the
projector in Eq.~\eqref{eq:vacuum-projector} is not automatically a unique
rank-one projector on the full Hilbert space.

For example, if
\begin{equation}
\mathcal H_{\rm phys}
=
\bigoplus_\alpha \mathcal H_\alpha ,
\end{equation}
and the asymptotic algebra acts sectorwise,
\begin{equation}
\mathcal A_\infty:\mathcal H_\alpha\to\mathcal H_\alpha ,
\end{equation}
then even a bounded-below Hamiltonian may have a sector-degenerate vacuum
space,
\begin{equation}
\ker H_{\rm bdry}
=
\bigoplus_\alpha {\rm span}\{|0_\alpha\rangle\}.
\end{equation}
The low-temperature limit gives a projector onto the sum of vacuum sectors,
\begin{equation}
\lim_{\beta\to\infty}e^{-\beta(H_{\rm bdry}-E_0)}
=
\sum_\alpha |0_\alpha\rangle\langle0_\alpha| ,
\end{equation}
not a unique projector $|0\rangle\langle0|$ on the full Hilbert space. The
sector projectors
\begin{equation}
P_{0,\alpha}=|0_\alpha\rangle\langle0_\alpha|
\end{equation}
then support completeness only within each $\mathcal H_\alpha$:
\begin{equation}
\overline{
\mathcal A_\infty^{(\alpha)}|0_\alpha\rangle
}
=
\mathcal H_\alpha .
\end{equation}

Thus the HoI argument is naturally representation-wise under these
assumptions. The question is whether the chosen representation is further
split by baby-universe superselection data, or whether a complete unitary
theory removes such data. This is the question to which we now turn.

\section{Topology sums and the sector obstruction}
\label{sec:topology-jt}

\subsection{Topology sums and sector structure}
\label{sec:topology-jtA}

A gravitational path integral need not be restricted to a fixed topology. One
may formally write
\begin{equation}
Z
=
\sum_{\rm topologies}
\int \mathcal D g\, e^{-I[g]} .
\label{eq:path-integral}
\end{equation}
Euclidean wormholes and disconnected geometries can then contribute to the
path integral. Such contributions have long been associated with baby
universes and $\alpha$-parameters \cite{Coleman:1988cy,Giddings:1988wv}, and
they have reappeared in modern discussions of JT gravity, ensemble averaging,
and replica wormholes \cite{Saad:2019lba,Marolf:2020xie,Blommaert:2020seb}.

The correct interpretation of such topology sums is not unique. We will use
the following trichotomy. First, the topology-summed path integral may produce
genuine central data, represented by a baby-universe Hilbert space and
$\alpha$-sectors. This is the canonical interpretation developed by Marolf and
Maxfield: one constructs a baby-universe Hilbert space $\mathcal H_{\rm BU}$
by cutting open compact components of the gravitational path integral;
boundary-insertion operators $\widehat Z[J]$ act on $\mathcal H_{\rm BU}$,
mutually commute, and are simultaneously diagonalized by $\alpha$-states. In a
fixed $\alpha$-state, amplitudes factorize; in a general state, such as the
Hartle--Hawking state, one obtains an ensemble over $\alpha$-sectors
\cite{Marolf:2020xie}. Second, the nonperturbative completion may remove this
structure: the relevant wormholes may be excluded, their contributions may
cancel, the integration contour may be constrained, or a quotient by null
states/gauge redundancies may collapse $\mathcal H_{\rm BU}$ to a single
state. A sharp version of this second branch is the Baby Universe Hypothesis
of McNamara and Vafa,
\begin{equation}
\dim \mathcal H_{\rm BU}=1,
\end{equation}
which they argue should hold in consistent unitary quantum gravity in $d>3$
\cite{McNamara:2020uza}. Third, the topology-summed path integral may describe
an effective or open gravitational subsystem, rather than a standalone closed
unitary theory. Low-dimensional ensemble examples, including JT gravity, and
AdS systems coupled to an external bath are natural places where this
possibility can arise.

This trichotomy separates two notions that are sometimes conflated. In a
bottom-up semiclassical saddle expansion, non-factorizing wormhole saddles are
not excluded by local semiclassical equations of motion alone. Their absence
in a closed unitary theory must come from global input in the definition of the
path integral: a top-down holographic constraint, a choice of contour,
nontrivial cancellations, or a null-state quotient. Standard AdS/CFT may well
realize this second branch. Starting from the exact boundary CFT, one expects
factorization and unitarity, and these properties can constrain the allowed
bulk topology sum \cite{Maldacena:1997re,McNamara:2020uza}. But this is a
constraint supplied by the nonperturbative definition of the theory, not a
purely local bottom-up semiclassical exclusion principle.

This observation is directly relevant for HoI. The HoI argument is often
motivated using semiclassical gravitational constraints, the boundary nature
of the Hamiltonian, and vacuum-sector cyclicity. If the topology-sum sector
data are removed by the exact theory, then the obstruction discussed below is
absent. But if the effective or exact gravitational description retains
nontrivial $\alpha$-sector data, then the same representation-wise HoI
reasoning gives completeness only after conditioning on $\alpha$. In that
case, promoting HoI to a statement on the full topology-summed Hilbert space
requires additional input.

The present paper focuses on this first branch as a conditional diagnostic.
If topology sums survive as genuine baby-universe sector data, the
Hilbert-space structure is enlarged beyond a single parent-universe
representation. Instead of a single non-$\alpha$-decomposed Hilbert space, one
obtains
\begin{equation}
\mathcal H_{\rm phys}
=
\int^\oplus d\alpha\, \mathcal H_\alpha ,
\label{eq:alpha-decomp}
\end{equation}
where $\alpha$ labels baby-universe or topology-sector data. From the point of
view of a single asymptotic region, these labels behave as superselection
data. The asymptotic algebra acts within each sector,
\begin{equation}
\mathcal A_\infty:\mathcal H_\alpha\to \mathcal H_\alpha ,
\end{equation}
and does not generically contain inter-sector operators,
\begin{equation}
|\alpha\rangle\langle\beta|,
\qquad \alpha\neq\beta .
\end{equation}
Hence asymptotic completeness, if valid, is naturally formulated as the
$\alpha$-sectorwise statement in Eq.~\eqref{eq:sector-hoi}. If the theory
instead realizes the second or third branch, then either the obstruction is
absent because $\mathcal H_{\rm BU}$ is trivial, or the topology-summed path
integral should be interpreted as an effective/open-system description in
which HoI need not be expected to hold on the gravitational subsystem alone.

\begin{figure}[t]
    \centering
    \IfFileExists{topology_theta_vacuum.png}{%
    \includegraphics[width=\columnwidth]{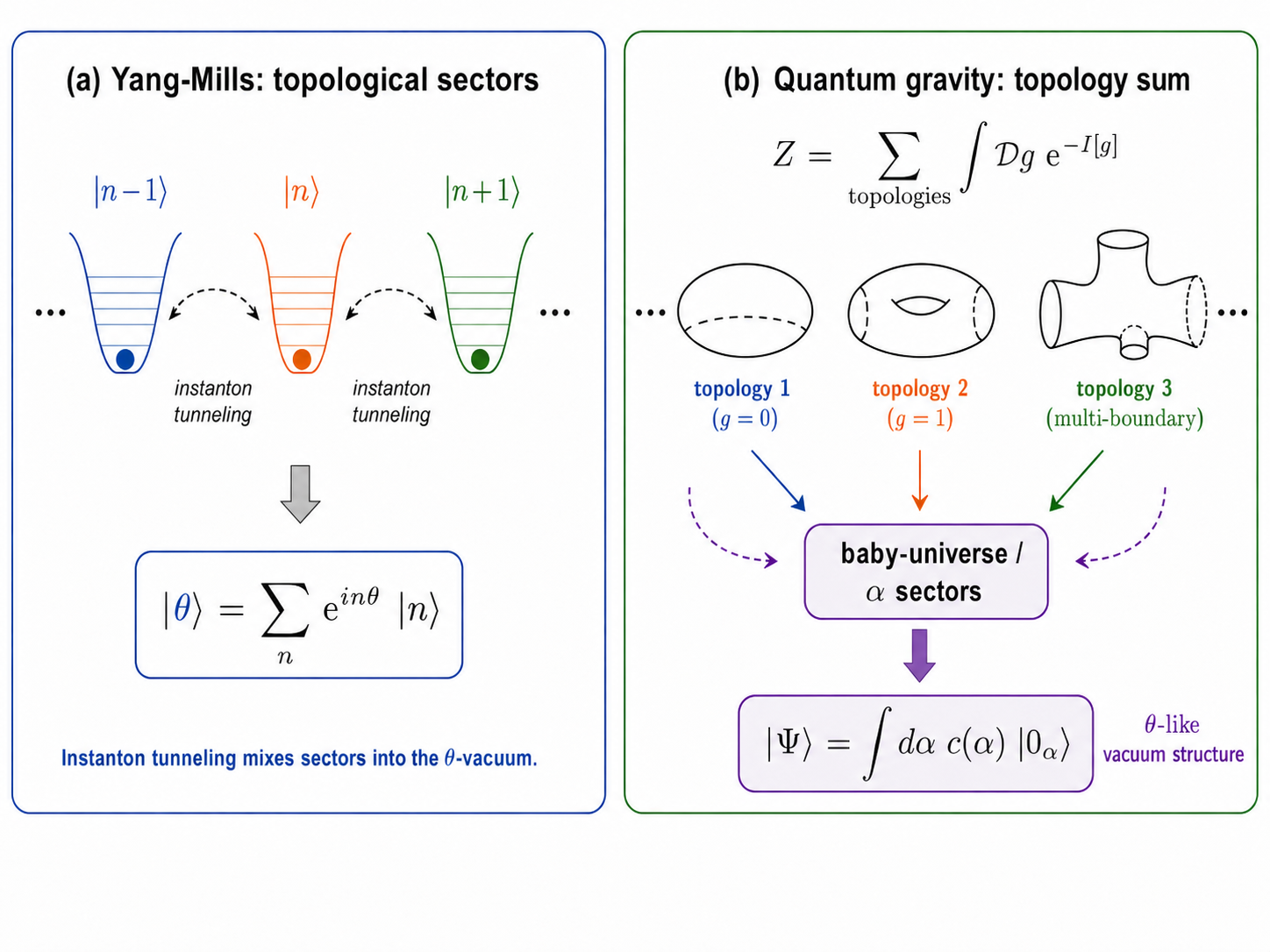}%
    }{%
    \fbox{\parbox{0.88\columnwidth}{\centering
    Schematic comparison of Yang--Mills topological sectors and topology
    sums in quantum gravity.}}%
    }
    \caption{Comparison between Yang--Mills topological sectors and topology
    sums in quantum gravity. \textbf{(a)} In Yang--Mills theory, instanton
    tunneling mixes distinct topological sectors $\lvert n\rangle$ into the
    $\theta$-vacuum, $\lvert\theta\rangle=\sum_n e^{in\theta}\lvert n\rangle$.
    \textbf{(b)} In topology-summed quantum gravity, different spacetime
    topologies can give rise to baby-universe or $\alpha$-sectors, leading to
    a $\theta$-like sector structure
    $\lvert\Psi\rangle=\int d\alpha\,c(\alpha)\lvert0_\alpha\rangle$.}
    \label{fig:theta-topology}
\end{figure}

This sectorwise structure is analogous in spirit to the familiar
topological-sector structure of Yang--Mills theory, although the physical
interpretation is different: in Yang--Mills theory instantons mix
topological sectors into a $\theta$-vacuum, whereas in topology-summed
gravity baby-universe data can appear as superselection labels for parent
universes. This analogy is summarized schematically in
Fig.~\ref{fig:theta-topology}.

\subsection{JT gravity as a diagnostic example}

Topology-summed JT gravity \cite{Jackiw:1984je,Teitelboim:1983ux} gives a concrete diagnostic example of this
structure. We emphasize that the appearance of baby-universe or
$\alpha$-sectors in JT gravity is not a new claim of this paper. It is an
established consequence of the interpretation of Euclidean wormholes and
matrix-integral averages in terms of baby-universe Hilbert spaces
\cite{Saad:2019lba,Marolf:2020xie,Blommaert:2020seb}. Our use of JT gravity is
diagnostic: we ask what this already-known sector structure implies for an
attempt to formulate HoI on a full topology-summed Hilbert space.

This use of JT gravity is not meant to claim that every complete unitary
quantum gravity has nontrivial baby-universe sectors. McNamara and Vafa have
argued that low-dimensional ensemble examples such as JT gravity should be
viewed as exceptional or effective descriptions, rather than as standalone
higher-dimensional complete quantum-gravitational systems satisfying the
usual swampland expectations \cite{McNamara:2020uza}. JT gravity is therefore
used here as a controlled diagnostic of the $\alpha$-sector branch, not as a
universal model of quantum gravity.

In JT gravity, the path integral includes geometries of arbitrary genus and
arbitrary numbers of asymptotic boundaries. The resulting amplitudes are
reproduced by matrix-integral averages \cite{Saad:2019lba,Blommaert:2020seb},
\begin{equation}
\langle Z(\beta_1)\cdots Z(\beta_n)\rangle .
\end{equation}
Euclidean wormholes generate connected contributions between disconnected
boundaries,
\begin{equation}
\langle Z(\beta_1)Z(\beta_2)\rangle_{\rm conn}\neq0,
\end{equation}
showing that the topology-summed gravitational path integral does not
factorize in the way expected of a single microscopic Hamiltonian with
independent copies.

Equivalently, the topology-summed JT path integral may be described using a
baby-universe Hilbert space. In a free baby-universe approximation one may
write schematically \cite{Blommaert:2020seb}
\begin{equation}
H(\phi)=H_0(\phi)+(a+a^\dagger)V(\phi),
\end{equation}
where $a,a^\dagger$ create and annihilate baby universes. Diagonalizing the
baby-universe operator,
\begin{equation}
(a+a^\dagger)|\alpha\rangle=\alpha|\alpha\rangle ,
\end{equation}
gives
\begin{equation}
H(\phi)|\alpha\rangle
=
\bigl(H_0(\phi)+\alpha V(\phi)\bigr)|\alpha\rangle .
\end{equation}
Thus different $\alpha$-states correspond to different effective parent
Hamiltonians,
\begin{equation}
H_\alpha=H_0+\alpha V .
\end{equation}
The sector decomposition is therefore not a pathology of the Hamiltonian; it
is the ordinary baby-universe interpretation of the topology-summed path
integral. In particular, the low-energy JT spectral density is bounded below
in the standard convention,
\begin{equation}
\rho_0(E)\propto \sinh(2\pi\sqrt{E})\,\theta(E).
\end{equation}

The Hartle--Hawking or no-boundary state prepared by the topology-summed path
integral may be written schematically as
\begin{equation}
|\Psi_{\rm HH}\rangle
=
\int d\alpha\, c(\alpha)|0_\alpha\rangle .
\label{eq:hh}
\end{equation}
A single-boundary asymptotic algebra acts within each $\alpha$-sector. Acting
on $|\Psi_{\rm HH}\rangle$ therefore gives states of the form
\begin{equation}
\mathcal A_\partial|\Psi_{\rm HH}\rangle
=
\int d\alpha\, c(\alpha)\,\mathcal H_\alpha ,
\end{equation}
with the same sector weights fixed by the baby-universe state. It does not
generically produce arbitrary states
\begin{equation}
|\Phi\rangle
=
\int d\alpha\, d(\alpha)|\phi_\alpha\rangle
\end{equation}
with independent coefficients $d(\alpha)$. Thus
\begin{equation}
\overline{\mathcal A_\partial|\Psi_{\rm HH}\rangle}
\neq
\mathcal H_{\rm full}
\end{equation}
unless additional assumptions are made about the accessibility of sector
projectors or inter-sector operators.

This is the role of topology-summed JT gravity in the present paper. It
supplies an already-known example in which a topology-summed path integral
naturally leads to sectorwise Hilbert spaces. We now isolate the corresponding
algebraic obstruction to a sector-independent HoI statement on the full topology-summed Hilbert space.

\subsection{Algebraic obstruction in the sector branch}
In the first branch of the trichotomy, the obstruction can be summarized in a simple algebraic way. Suppose
that the topology-summed Hilbert space admits a central decomposition
\begin{equation}
\mathcal H_{\rm full}
=
\int^\oplus d\alpha\,\mathcal H_\alpha ,
\end{equation}
where $\alpha$ labels baby-universe or topology-sector data. If asymptotic
observables associated with a single parent universe preserve the
$\alpha$-sector, then they are decomposable operators,
\begin{equation}
\mathcal A_\infty
\subset
\int^\oplus d\alpha\,\mathcal B(\mathcal H_\alpha).
\end{equation}
Equivalently, they commute with the central algebra generated by the
baby-universe label. For any bounded function $f(\alpha)$, define
\begin{equation}
Z_f
=
\int^\oplus d\alpha\, f(\alpha)\,\mathbf 1_\alpha .
\end{equation}
Then
\begin{equation}
[A,Z_f]=0,
\qquad
A\in \mathcal A_\infty .
\end{equation}

Now consider a rank-one operator connecting two distinct sectors,
\begin{equation}
T_{\alpha\beta}
=
|\psi_\alpha\rangle\langle \chi_\beta|,
\qquad
\alpha\neq \beta ,
\end{equation}
with $|\psi_\alpha\rangle\in\mathcal H_\alpha$ and
$|\chi_\beta\rangle\in\mathcal H_\beta$. Its commutator with $Z_f$ is
\begin{equation}
[Z_f,T_{\alpha\beta}]
=
\bigl(f(\alpha)-f(\beta)\bigr)T_{\alpha\beta}.
\end{equation}
For a function $f$ with $f(\alpha)\neq f(\beta)$, this commutator is nonzero.
Therefore $T_{\alpha\beta}$ cannot belong to a sector-preserving asymptotic
algebra:
\begin{equation}
T_{\alpha\beta}\notin \mathcal A_\infty .
\end{equation}
But such inter-sector rank-one operators are elements of
$\mathcal B(\mathcal H_{\rm full})$. Hence a sector-preserving asymptotic
algebra cannot be globally complete:
\begin{equation}
\mathcal A_\infty
\neq
\mathcal B(\mathcal H_{\rm full}) .
\end{equation}

Thus, within the sector branch, an extension of HoI to the full
topology-summed Hilbert space requires at least one of the following
additional conditions: the $\alpha$-decomposition is absent, the theory
projects onto a single $\alpha$-sector, or the asymptotic algebra contains
operators that resolve or connect different $\alpha$-sectors. Without such
additional input, the natural conclusion is $\alpha$-sectorwise HoI,
\begin{equation}
\overline{\mathcal A_\infty^{(\alpha)}|0_\alpha\rangle}
=
\mathcal H_\alpha ,
\end{equation}
rather than HoI on the full topology-summed Hilbert space $\mathcal H_{\rm full}$.

\section{Sectorwise HoI and horizon normalcy}
\label{sec:normalcy}

\subsection{The conceptual obstruction}

We now turn to the main point. HoI is used in black hole physics to challenge
the AMPS factorization assumption. Semiclassical EFT suggests an approximate
factorization
\begin{equation}
\mathcal H_{\rm full}
\simeq
\mathcal H_{\rm ext}\otimes \mathcal H_{\rm int}
\end{equation}
near a black hole horizon. AMPS uses this approximate factorization, together
with unitarity and effective field theory, to argue for a conflict between
late-time unitarity and horizon smoothness. The HoI argument attempts to evade this
conflict by claiming that the exterior/asymptotic algebra is already complete
in the exact theory. If so, the interior is not an independent tensor factor.

Within a fixed $\alpha$-sector, this logic may go through:
\begin{equation}
\overline{
\mathcal A_\infty^{(\alpha)}|0_\alpha\rangle
}
=
\mathcal H_\alpha .
\end{equation}
The asymptotic observer may then attempt to deny an independent
interior-exterior factorization inside $\mathcal H_\alpha$. But in the full topology-summed
description one has, instead,
\begin{equation}
\mathcal H_{\rm full}
=
\int^\oplus d\alpha\,\mathcal H_\alpha ,
\end{equation}
and the corresponding completeness statement on the full topology-summed
Hilbert space,
\begin{equation}
\overline{\mathcal A_\infty|0\rangle}
=
\mathcal H_{\rm full},
\end{equation}
does not follow without additional assumptions. Thus the obstruction to
factorization becomes representation-dependent: it holds after choosing the
sector in which HoI is asserted.

This is not merely a technical weakening of HoI. It changes the status of
horizon smoothness. A local infalling observer is normally expected to
encounter a smooth horizon whenever the local semiclassical geometry and
state are sufficiently close to the appropriate black hole vacuum. This
expectation is local and generally covariant. It should not depend on a
global baby-universe label invisible to the local observer.

But if the HoI-based resolution of AMPS works only after choosing an $\alpha$-sector,
then the absence of a firewall is no longer a purely local consequence of the
semiclassical geometry. Instead,
\begin{equation}
\text{HoI-normalcy}
\ \Longrightarrow\
\alpha\text{-fixed or sector-aligned}.
\label{eq:smooth-sector}
\end{equation}
Thus the criterion for HoI-based normalcy becomes conditional on the chosen
sector or on a sector-independent alignment of the interior reconstruction.

\subsection{A Bell-pair diagnostic}

The preceding argument is a statement about global operator algebras. To see
how it can show up in the near-horizon data used in the AMPS argument, we now
make the obstruction quantitative in a minimal model of the
AMPS near-horizon degrees of freedom. In the AMPS argument, a late outgoing
Hawking mode $b$ must be entangled with its interior partner $\tilde b$ in
order for the horizon to be smooth, while unitarity tends to correlate $b$
with the early radiation. Since only a finite number of states of a single
wavepacket are relevant in an effective description, we model $b$ and
$\tilde b$ as $d$-dimensional systems. This finite-dimensional model is not
meant to capture the full field theory; it isolates the algebraic question of
whether the Bell-pair structure required for smoothness can be chosen
independently of the baby-universe sector.

The tensor product used in this diagnostic is the approximate organization of
near-horizon degrees of freedom supplied by semiclassical effective field
theory; it is not an exact factorization of the complete Hilbert space
$\mathcal H_\alpha$. More precisely, for each sector we use a local code space
\begin{equation}
\mathcal H_{{\rm code},\alpha}
\simeq
\mathcal H_b\otimes\mathcal H_{\tilde b}^{\rm EFT},\label{eq:local_code_space}
\end{equation}
together with an isometric embedding
\begin{equation}
W_\alpha:
\mathcal H_{{\rm code},\alpha}
\longrightarrow
\mathcal H_\alpha .
\label{eq:code-embedding}
\end{equation}
Sectorwise HoI denies that $\mathcal H_{\tilde b}^{\rm EFT}$ is an independent
\emph{exact} tensor factor of $\mathcal H_\alpha$; it remains compatible with
using this approximate code-space factorization to formulate the local
smoothness condition.

In a fixed $\alpha$-sector, horizon smoothness requires $b$ and its interior
partner to be in an Unruh-like maximally entangled state. We write a reference
smooth state as
\begin{equation}
\ket{\Phi}
=
\frac{1}{\sqrt d}
\sum_{i=1}^d
\ket{i}_b\ket{i}_{\tilde b}.
\end{equation}
If the HoI reconstruction of the interior is sector-dependent, then the
identification of the interior partner may differ from sector to sector. We
parametrize this dependence by a unitary $U_\alpha$ acting on the partner
Hilbert space:
\begin{equation}
\ket{\Phi_\alpha}
=
(\mathbf 1_b\otimes U_\alpha)\ket{\Phi}.
\label{eq:sector-bell}
\end{equation}
Physically, $U_\alpha$ represents the $\alpha$-dependent mirror map,
gravitational dressing, or reconstruction of the interior operator within the
chosen sector. Thus each fixed $\alpha$-sector may have a perfectly smooth
horizon, but the map identifying which operator is the interior partner need
not be the same in different sectors.

The vector $\ket{\Phi_\alpha}$ belongs to the auxiliary code space. The
corresponding physical smooth state is
\begin{equation}
\ket{\Psi^{\rm smooth}_\alpha}
=
W_\alpha\ket{\Phi_\alpha}
\in\mathcal H_\alpha .
\label{eq:physical-sector-bell}
\end{equation}
Accordingly, Eqs.~\eqref{eq:local_code_space}--\eqref{eq:physical-sector-bell} do not assume an exact decomposition
$\mathcal H_\alpha=\mathcal H_b\otimes\mathcal H_{\tilde b}$. They only encode
the semiclassical near-horizon subsystem on which the Bell-pair test is
performed.

A sector-blind description does not condition on $\alpha$. On
the exact sector-preserving algebra, the corresponding state is block
diagonal,
\begin{equation}
\bar\rho_{\rm phys}
=
\int^\oplus d\alpha\,p(\alpha)\,
W_\alpha\ket{\Phi_\alpha}\bra{\Phi_\alpha}W_\alpha^\dagger .
\label{eq:sector-averaged-physical-rho}
\end{equation}
To compare the conditional near-horizon reconstructions, we choose a common
auxiliary code-space convention and pull each block back along $W_\alpha$.
The density matrix used in the diagnostic is therefore
\begin{equation}
\bar\rho_{b\tilde b}
=
\int d\alpha\,p(\alpha)\,
\ket{\Phi_\alpha}\bra{\Phi_\alpha}.
\label{eq:sector-averaged-rho}
\end{equation}
It is an averaged code-space diagnostic, not a density matrix on a single
exact sector-independent interior--exterior tensor product.
To test whether this state is smooth with respect to a single
sector-independent interior reconstruction, compare it with a candidate
smooth Bell state
\begin{equation}
\ket{\Phi_V}
=
(\mathbf 1_b\otimes V)\ket{\Phi},
\end{equation}
where $V$ is a fixed unitary, independent of $\alpha$. The Bell-pair fidelity
is
\begin{align}
F(V)
&=
\bra{\Phi_V}\bar\rho_{b\tilde b}\ket{\Phi_V}
\nonumber\\
&=
\int d\alpha\,p(\alpha)\,
\left|
\bra{\Phi}
(\mathbf 1_b\otimes V^\dagger U_\alpha)
\ket{\Phi}
\right|^2
\nonumber\\
&=
\int d\alpha\,p(\alpha)\,
\frac{\left|\mathrm{Tr}(V^\dagger U_\alpha)\right|^2}{d^2}.
\label{eq:bell-fidelity}
\end{align}
In the last step we used the standard identity
\begin{equation}
\bra{\Phi}(\mathbf 1\otimes W)\ket{\Phi}
=
\frac{1}{d}\mathrm{Tr}\,W .
\end{equation}
Since
\begin{equation}
\left|\mathrm{Tr}(V^\dagger U_\alpha)\right|
\leq d ,
\end{equation}
we have $F(V)\leq 1$. Equality holds only if
\begin{equation}
V^\dagger U_\alpha
=
e^{i\varphi(\alpha)}\mathbf 1
\end{equation}
for $p(\alpha)$-almost every $\alpha$. Thus a sector-independent smooth
interior exists only if all sector-dependent interior reconstructions
$U_\alpha$ agree up to phases. A common change of the auxiliary
code-space convention can be absorbed into $V$; the criterion therefore tests
only the relative $\alpha$-dependence that cannot be removed by one
sector-independent choice. Otherwise,
\begin{equation}
\max_V F(V)<1 .
\end{equation}
This gives a quantitative version of the obstruction: sectorwise smoothness
does not imply the existence of a single sector-independent smooth horizon.

The loss can be maximal in a limiting case. Suppose that the sector average
induces a completely depolarizing channel on the interior partner,
\begin{equation}
\mathcal E(X)
=
\int d\alpha\,p(\alpha)\,
U_\alpha X U_\alpha^\dagger
=
\frac{\mathrm{Tr}\,X}{d}\,\mathbf 1
\end{equation}
for every operator $X$ on the partner Hilbert space. Equivalently, the
sector-dependent reconstructions form a unitary one-design on the partner
system. Such a situation would arise if the baby-universe sector effectively
samples a complete set of mutually dephasing interior reconstructions, for
example through sufficiently random $\alpha$-dependent phases and mixings in
the mirror map, or in an ensemble interpretation where different microscopic
Hamiltonians induce uncorrelated interior identifications. The one-design
assumption is therefore not an assumption about generic JT dynamics; it is a
controlled diagnostic of the maximal possible loss of a sector-independent
smooth interior.

Under this assumption,
\begin{equation}
\bar\rho_{b\tilde b}
=
(\mathbf 1_b\otimes \mathcal E)\bigl(|\Phi\rangle\langle\Phi|\bigr)
=
\frac{\mathbf 1_b}{d}\otimes
\frac{\mathbf 1_{\tilde b}}{d}.
\end{equation}
Consequently,
\begin{equation}
I(b:\tilde b)_{\bar\rho}=0,
\qquad
F(V)=\frac{1}{d^2}
\end{equation}
for every sector-independent choice of $V$. Thus every fixed $\alpha$-sector
may contain a perfectly smooth near-horizon Bell pair, while the
sector-averaged description contains no Bell-pair correlation at all.

This calculation does not claim that topology-summed JT gravity literally
produces a one-design distribution of interior reconstructions. Rather, it
shows that sector-independent normalcy requires an additional alignment condition, or
else explicit conditioning on the sector label: the sector-dependent maps
$U_\alpha$ must agree up to phases, or the sector label must be accessible so
that one can condition on $\alpha$. Without such an alignment or conditioning,
smoothness is a statement within each sector, not a sector-independent
property of the topology-summed state.

\subsection{Relation to general covariance}

The preceding discussion sharpens the relation to Bousso's argument in
Ref.~\cite{Bousso:2025udh}. Bousso defines horizon normalcy as the approximate
validity of semiclassical gravity and effective field theory in a neighborhood
of an event \(E\) at which an observer approaches or crosses the black hole
horizon. In the AMPS setup, normalcy requires the outgoing near-horizon mode
\(b\) to be entangled with its interior partner \(\tilde b\), so that the
near-horizon state is locally close to the Unruh vacuum. On the other hand,
unitarity together with effective propagation to infinity implies that the
same outgoing mode \(b\) is purified by a subsystem \(b_R\) distilled from the
early Hawking radiation. The simultaneous requirements
\begin{equation}
S_{b\tilde b}\ll 1,
\qquad
S_{b b_R}\ll 1,
\qquad
S_b=O(1),
\end{equation}
are incompatible with strong subadditivity. Indeed, strong subadditivity implies \cite{Mathur:2009hf}
\begin{equation}
S_{b\tilde b}+S_{bb_R}\geq S_b ,
\end{equation}
which is incompatible with the simultaneous purity of \(b\tilde b\) and
\(bb_R\) when \(S_b=O(1)\). Thus, whenever the exact theory
admits the semiclassical factorization needed to regard \(b\), \(\tilde b\),
and \(b_R\) as independent systems, the AMPS monogamy obstruction implies a
failure of horizon normalcy.

The HoI proposal attempts to avoid this conclusion by denying the relevant factorization.

In a fixed \(\alpha\)-sector this is the intended mechanism: if
\begin{equation}
\overline{\mathcal A_\infty^{(\alpha)}|0_\alpha\rangle}
=
\mathcal H_\alpha ,
\end{equation}
then the asymptotic algebra is complete in that representation, and the
interior partner \(\tilde b\) is not an independent tensor factor. The outgoing
mode \(b\) may then be described as having a smooth interior partner without
violating unitarity, because the AMPS factorization premise has been removed
inside \(\mathcal H_\alpha\).

The obstruction arises when one asks for a sector-independent statement in
the full topology-summed theory. The unprojected state has the schematic form
\begin{equation}
|\Psi\rangle
=
\int d\alpha\,c(\alpha)\,|\Psi_\alpha\rangle .
\end{equation}
If one evolves a global semiclassical slice in this full theory without first
conditioning on a particular \(\alpha\)-sector, then the late Hawking mode
\(b\) must still be correlated with the exterior radiation in order to
preserve unitarity. But the sector-independent HoI statement needed to deny
interior-exterior factorization has not been established on
\(\mathcal H_{\rm full}\). In that description, the AMPS logic can reappear:
\begin{align}
&b \text{ purified by } b_R
\quad
\Longrightarrow
\quad
b \text{ cannot also be maximally} \notag\\&\hspace{4.3cm}\text{purified by $\tilde b$}.
\end{align}
Thus the same near-horizon mode is treated differently depending on whether
one first projects onto a fixed \(\alpha\)-sector or evolves the full
topology-summed state.

This is the precise sense in which the HoI-based criterion for horizon
normalcy becomes representation-dependent. In a chosen \(\alpha\)-sector, the HoI argument can obstruct
factorization and allow the infalling observer to see a normal horizon. In the
unprojected topology-summed description, the algebraic reason for rejecting
factorization is absent unless additional sector-aligning or sector-resolving
data are supplied. The criterion for normalcy is therefore not simply a local
statement about the semiclassical geometry near \(E\); it depends on whether
the global quantum-gravitational state has been conditioned on a particular
baby-universe sector.

This has the same logical structure as Bousso's covariance objection. In
Bousso's discussion, if horizon normalcy depends on the state of distant
Hawking radiation, then one must either select a preferred global time slice
on which that radiation state is evaluated, or allow the normalcy of \(E\) to
depend on the infinite future of the exterior. In the present setting, the
role of the distant radiation state is played by the global sector data of
the topology-summed Hilbert space. If horizon normalcy depends on whether the
description is sector-projected or sector-blind, then the smoothness of the
horizon is not determined by local semiclassical data alone.

Therefore, the conclusion is not that a firewall must form in every sector.
Rather, the conclusion is that the HoI-based avoidance of AMPS is not
automatically a generally covariant, state-independent statement once topology
sums are included. The HoI argument may still be formulated within a fixed
sector, but without a sector-independent alignment of the interior reconstruction, or an
asymptotic mechanism that resolves the \(\alpha\)-sector, the absence of a
firewall becomes conditional on a global sector or representation choice.

\begin{figure}[t]
    \centering
    \IfFileExists{normalcy.png}{%
    \includegraphics[width=0.98\columnwidth]{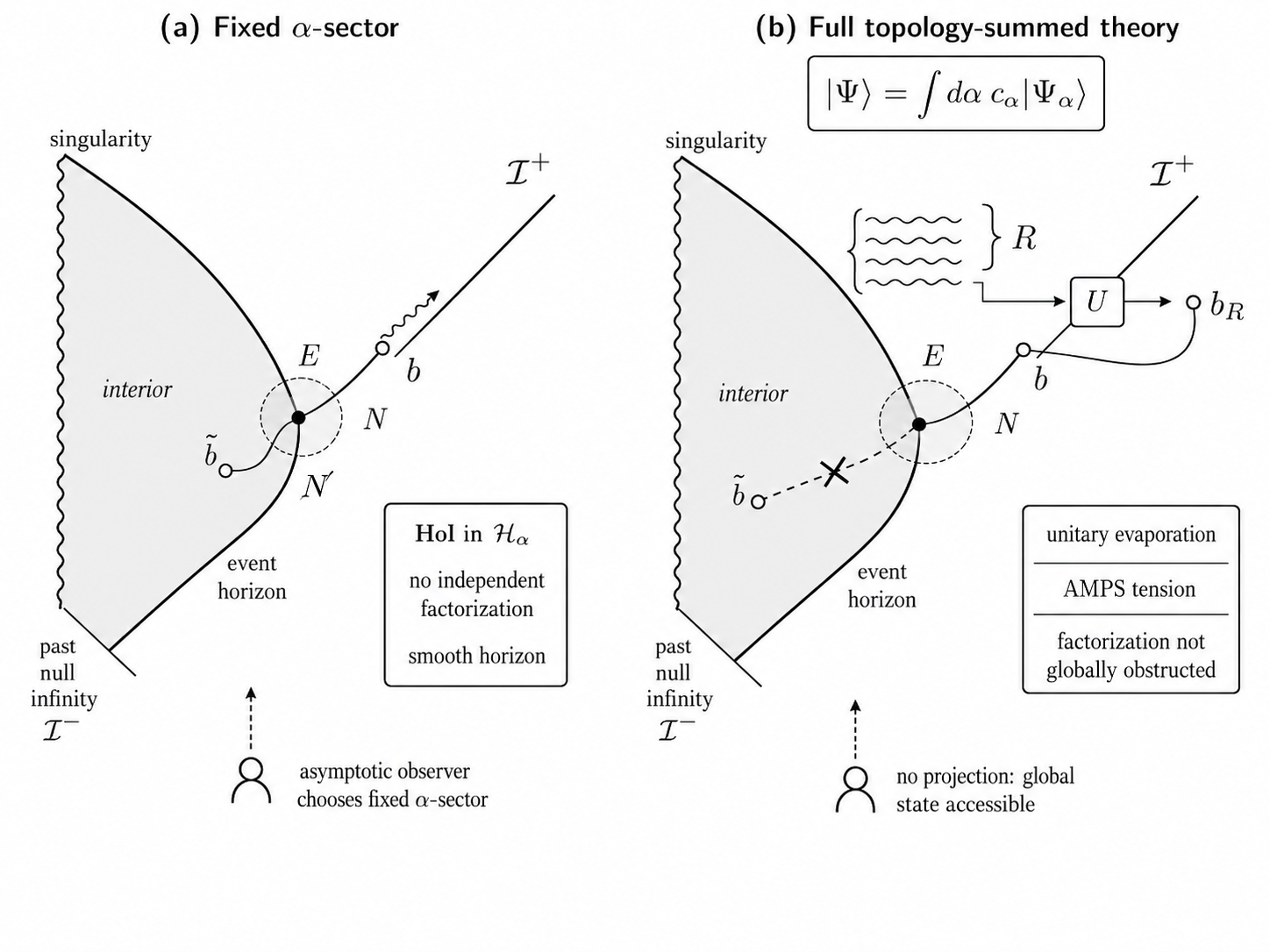}%
    }{%
    \fbox{\parbox{0.88\columnwidth}{\centering
    Schematic comparison between an $\alpha$-conditioned description and an
    unconditioned topology-summed description of horizon normalcy.}}%
    }
    \caption{
    Schematic comparison between sectorwise HoI and the unprojected
    topology-summed description.
    \textbf{(a)} In a fixed baby-universe or $\alpha$-sector, the HoI argument
    may render the asymptotic algebra complete on $\mathcal H_\alpha$. The
    near-horizon outgoing mode $b$ can then be identified with an interior
    partner $\tilde b$ without invoking an independent interior tensor factor,
    allowing the infalling observer crossing the event $E$ to experience a
    smooth horizon.
    \textbf{(b)} In the full topology-summed state,
    $|\Psi\rangle=\int d\alpha\,c_\alpha|\Psi_\alpha\rangle$, no fixed
    $\alpha$-sector has been selected. The late mode $b$ is correlated with the
    exterior Hawking radiation $R$, from which a purifier $b_R$ may be
    distilled by a decoding operation $U$. Since the sector-independent HoI obstruction to
    factorization is not established on the unprojected Hilbert space, the
    AMPS monogamy tension can reappear: $b$ cannot be simultaneously purified
    by both $b_R$ and the interior partner $\tilde b$. Thus the normalcy of the
    same local horizon-crossing event $E$ becomes conditional on the global
    sector or representation choice.
    }
    \label{fig:sectorwise-normalcy}
\end{figure}

This is shown in Fig.~\ref{fig:sectorwise-normalcy}.

\section{Discussion}
\label{sec:discussion}

In this work, we examined the relation between HoI and topology-summed
Hilbert-space structures. The first lesson is terminological but important.
The usual HoI argument is already formulated within a chosen representation:
one fixes the relevant vacuum or boundary-condition sector and then asks
whether the asymptotic algebra is complete on that Hilbert space,
\begin{equation}
\overline{\mathcal A_\infty |0\rangle}
=
\mathcal H_{\rm vac} .
\end{equation}
Our question was whether this representation is further decomposed by
baby-universe or $\alpha$-sector data. If such data are present, then the
natural HoI statement becomes
\begin{equation}
\overline{\mathcal A_\infty^{(\alpha)}|0_\alpha\rangle}
=
\mathcal H_\alpha ,
\end{equation}
and one must still ask whether this can be promoted to a statement on the
full topology-summed Hilbert space.

Our analysis is conditional on the $\alpha$-sector branch of topology-summed
quantum gravity. Marolf and Maxfield provide the canonical Hilbert-space
framework for this branch: spacetime wormholes lead to a baby-universe
Hilbert space, commuting boundary-insertion operators, and $\alpha$-states
that define superselection sectors for the asymptotic observables
\cite{Marolf:2020xie}. In a fixed $\alpha$-state, amplitudes factorize; in a
general baby-universe state, one obtains an ensemble. If this structure
survives, then
\begin{equation}
\mathcal H_{\rm full}
=
\int^\oplus d\alpha\,\mathcal H_\alpha,
\end{equation}
and a sector-preserving asymptotic algebra cannot generate arbitrary
inter-sector rank-one operators. Hence completeness in each $\alpha$-sector
need not imply completeness on $\mathcal H_{\rm full}$:
\begin{equation}
\overline{\mathcal A_\infty^{(\alpha)}|0_\alpha\rangle}
=
\mathcal H_\alpha
\qquad
\not\Rightarrow
\qquad
\overline{\mathcal A_\infty|0\rangle}
=
\mathcal H_{\rm full} .
\end{equation}

However, this is not a claim that standard unitary AdS/CFT contains genuine
baby-universe $\alpha$-sectors. The same Marolf--Maxfield framework contains
the factorizing special case: if the baby-universe Hilbert space is
one-dimensional, the Hartle--Hawking state is the unique $\alpha$-state and no
ensemble is present. McNamara and Vafa have argued for precisely such a
collapse in consistent unitary quantum gravity in $d>3$, formulating the Baby
Universe Hypothesis
\begin{equation}
\dim \mathcal H_{\rm BU}=1
\end{equation}
as a swampland condition \cite{McNamara:2020uza}. If this hypothesis, or an
equivalent top-down AdS/CFT constraint, holds in the exact theory, then the
specific $\alpha$-sector obstruction discussed in this paper is absent and
HoI may remain complete in the relevant representation.

The remaining point is that this absence is itself part of the
nonperturbative definition of the theory. From a bottom-up semiclassical
saddle expansion, non-factorizing wormholes are not excluded by local
semiclassical equations alone. Their disappearance in a closed unitary theory
requires additional global input: the exact CFT definition, a restricted
contour, cancellations among topologies, or a null-state/gauge-redundancy
quotient. Thus the topology-sum question and the HoI question should be
understood as being decided by the same nonperturbative Hilbert-space
structure. If the completion collapses $\mathcal H_{\rm BU}$ to one dimension,
there is no $\alpha$-sector obstruction. If an effective or exact
gravitational description retains nontrivial $\alpha$-sector or ensemble data,
then HoI is naturally $\alpha$-sectorwise.

This is the sense in which the result is in tension with the generally
covariant notion of horizon normalcy emphasized by Bousso
\cite{Bousso:2025udh}. The issue is not that a firewall must form in every
sector. Rather, in the $\alpha$-sector branch, the existence or absence of a
firewall is no longer determined purely by local semiclassical geometry near
the horizon-crossing event. It depends on whether one has conditioned on a
global baby-universe sector, or supplied additional nonperturbative data that
makes the sector dependence irrelevant.

The final conclusion is therefore conditional. In a closed unitary theory
satisfying the Baby Universe Hypothesis, or otherwise excluding nontrivial
$\alpha$-sectors, the obstruction discussed here collapses. In a gravitational
description where $\alpha$-sector or ensemble data survive, a sectorwise HoI
statement is not enough to establish a state-independent and generally
covariant resolution of the firewall problem. Such a conclusion requires
additional input specifying whether the baby-universe sector structure is
absent, projected onto a single sector, asymptotically resolvable, or aligned
in the interior reconstruction.

\section*{Acknowledgments}
I thank Hao Geng for helpful comments and discussions on the scope of the manuscript and on the relation between holography of information, topology sums, and baby-universe sectors.

\bibliography{bib}
\bibliographystyle{utphys1}

\end{document}